# Schwarzschild-Couder telescope for the Cherenkov Telescope Array: Development of the Optical System

J. ROUSSELLE[1], V. CONNAUGHTON[2], M. ERRANDO[3], T. B. HUMENSKY[4], R. MUKHERJEE[3], D. NIETO[4], A. OKUMURA[5,6], V. V. VASSILIEV[1] FOR THE CTA CONSORTIUM.

[1] *University of California Los Angeles*
[2] *University of Alabama in Huntsville*
[3] *Barnard College, Columbia University*
[4] *Columbia University*
[5] *Nagoya University*
[6] *University of Leicester*

*rousselle@astro.ucla.edu*

**Abstract:** The Cherenkov Telescope Array (CTA) is the next generation ground-based observatory for very high-energy (E>100 GeV) gamma-ray astronomy. It will integrate several tens of imaging atmospheric Cherenkov telescopes (IACTs) with different apertures into a single astronomical instrument. The US part of the CTA collaboration has proposed and is developing a novel IACT design with a Schwarzschild-Couder (SC) aplanatic two-mirror optical system. In comparison with the traditional single mirror Davies-Cotton IACT the SC telescope, by design, can accommodate a wider field-of-view, with significantly improved imaging resolution. In addition, the reduced plate scale of an SC telescope makes it compatible with highly integrated cameras assembled from silicon photo multipliers. In this submission we report on the status of the development of the SC optical system, which is part of the effort to construct a full-scale prototype telescope of this type at the Fred Lawrence Whipple Observatory in southern Arizona.

## 1 Introduction

During the last decade, the scientific achievements of H.E.S.S., MAGIC and VERITAS observatories proved the technical feasibility and broad scientific value of the Imaging Atmospheric Cherenkov Telescopes (IACTs) in the very high energy (VHE) astronomy domain. The scientific community is now developing the Cherenkov Telescope Array (CTA)[1], the next generation IACT array made of several tens of telescopes. The US part of the CTA consortium will contribute to this international project leading the development of a mid-size Schwarzschild-Couder Telescope (SCT), a novel two mirrors optical design for IACT, with an aperture of 9.6 m. This design has the advantage of a wider field of view (8 degrees) and more compact plate scale compared to the traditional Davies-Cotton telescopes, allowing the use of high density photosensors, such as silicon photomultipliers (SiPMs). A full size prototype of such telescope is currently under development and will be built at the Fred Lawrence Whipple Observatory in Arizona within three years.

This paper reports on the current development status of the SCT optical system, including stray-light control, alignment tolerances as well as the main elements of the alignment system.

## 2 Overview of the optical system

The current IACTs, based on prime focus optical designs, are recognized for their performance and reliability. Nevertheless, they are affected by large comatic aberrations, only mitigated by a long focal length, leading to a large camera and heavy mechanical structure. In comparison, the secondary mirror of the SCT, demagnifying the image, allows a significant reduction of both focal length of the optical system and plate scale size, leading to an improved imaging resolution and larger field of view (see [2] and [3] for a detailed comparison).

The current SCT design developed for the mid-size CTA telescopes (Fig. 1 left) has a 5.6 m focal length for the optical system. It is composed of a 9.66 m aperture primary mirror with a 4.4 m central hole, a 5.4 m secondary mirror, and a 0.8 m diameter focal plane, corresponding to a field of view of 8 degrees. The figures of the two mirrors rely on the exact solution of Schwarzschild's equations, removing spherical and coma aberrations, while the curvature of the focal plane minimizes astigmatism [3].

Figure 1 (right) presents segmentation schemes, with the primary and secondary mirrors, respectively made of 48 and 24 panels distributed on two rings.

## 3 Stray light control

Given the significantly smaller angular size of the pixels on the focal plane (∼2x), noise from stray light photons becomes a significant contribution to the overall accidental rate. For the prototype SCT optical design, the estimated rate expected from stray light photons is ∼ 18 MHz, comparable to the rates expected from night sky background.

Stray light control elements can be added to the optical design to reduce the amount of stray light collected in the focal plane. These elements can be either obscuring baffles around the primary and secondary mirror, or the camera, or lenses that modify the angular acceptance of the camera pixels. Obscuring the central hole on the primary mirror

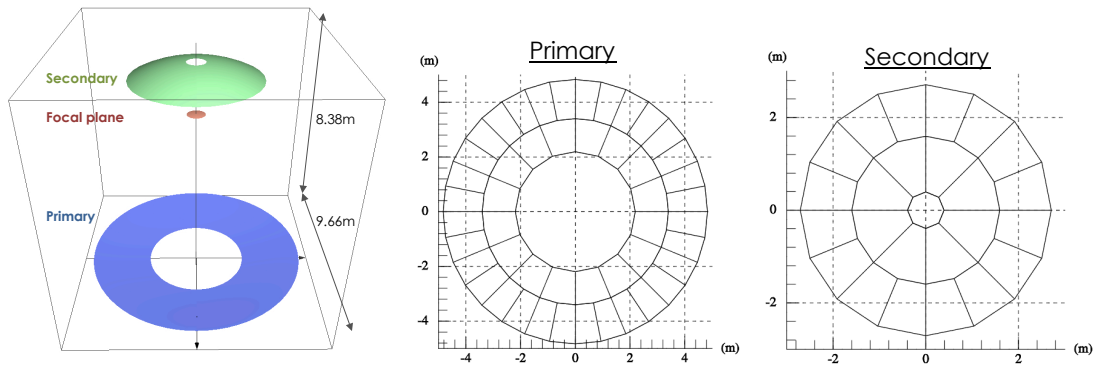

**Fig. 1**: Left: overview of the SCT optical system with an aperture of 9.66m. Middle and right are the favored segmentation schemes for the primary and secondary mirrors. They are made of 48 and 24 mirror panels, respectively.

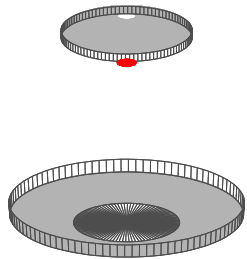

**Fig. 2**: Geometrical design of the baffles around and in the middle of the primary and secondary mirrors (presented in Figure 1), reducing the amount of stray light by 90%.

dish will also reduce the amount of stray light collected at the focal plane.

Ray-tracing simulations of the SCT optical design have shown that a combination of a 50 cm baffle around the primary mirror and a 30 cm baffle around the secondary mirror, together with obscuring the center of the primary dish, would reduce the amount of collected stray light by 90% (see Fig. 2). The advantage of this design is that a significant reduction in the amount of collected stray light can be achieved with only 1% reduction in collection area, coming from the shadowing, introduced by the baffle on the secondary mirror. Stray light control elements in or around the camera have been evaluated but will not be used due to the significantly higher shadowing that they introduce to the system.

## 4 Alignment tolerances

In order to develop the telescope mechanical structure and alignment system, one should first determine the alignment precision of the optical system needed for our specific application. The PSF size as well as the displacement of the PSF centroid position were used as quality factor to estimate these alignment tolerances, with the objective to keep the PSF size smaller than the size of a photodetector pixel on the camera, about 5 arcmin for the SCT design.

The tolerances have been divided in two parts, corresponding to the alignment of the whole mirrors on the optical axis (called global alignment) and the alignment of each mirror panels with respect to each other within the mirrors (called panel-to-panel alignment). This separation follow the different conceptual designs for global and panel-to-panel alignment systems presented in 5.1.

Extensive ray-tracing simulations were used to study independently each parameter having an impact on the PSF size and position. These parameters correspond to the rotations and translations of the mirrors as a whole and each individual mirror panel.

The objective was to identify the main parameters driving the degradation of the optical performance in order to develop the most efficient alignment system. To compare the sensitivity of the optical system to each of these parameters we estimated the amplitude of the transformation that must be applied to the mirrors in order to increase the PSF size to 1 arcmin on axis. Although this 1 arcmin value of the PSF size is arbitrary, it has been proven that the deformation of the optical system and the PSF size follow a linear relation around this value, allowing us to extrapolate the results presented here to different degradations.

Tables 1 and 2 show the resulting values for the global alignment and the panel-to-panel alignment, respectively for the primary and secondary mirrors. In the case of the panel-to-panel alignment, all of them have a random misalignment amplitude, following a Gaussian distribution centered on the ideal position.

The translation of the primary mirror along the optical axis has only a weak impact on the performance of the optical system (Table 1) with a translation of 17 mm needed to increase the PSF size on axis to 1 arcmin. Because this tolerance is easily achievable, we could choose to reduce this translation by an order of magnitude in order to allocate a bigger part of the error budget to much tighter constrains such as the rotation of the primary mirror panels.

Generally speaking, the optical system seems very sensitive to the rotations of the primary mirror panels and the translation of the secondary mirror panels. In addition, Table 1 shows a very relaxed tolerance on the tilt of the whole primary mirror, which is two orders of magnitude higher than the secondary mirror. Although this parameter has a very weak impact on the PSF size, it introduces a large off-set to the position of the PSF centroid. Thus the tolerance on the tilt of the whole primary mirror is defined by the accuracy of the PSF centroid position on the focal plane, rather than the PSF size.

The results of the ray-tracing simulations indicate the need for sub-millimeter and sub-milliradian precisions for both global and panel-to-panel alignments. Also, it has been proven that the tolerances presented here are almost independent of the segmentation scheme chosen for both mirrors.

| Primary mirror Global alignment | Value |
|---|---|
| Translation ⊥ to optical axis | 10 mm |
| Translation ∥ to optical axis | 17 mm |
| Tilt | 15 mrad |
| Panel alignment | Standard deviation |
| Translation ⊥ to optical axis | 2.2 mm |
| Translation ∥ to optical axis | 17 mm |
| Rotation around tangent axis | 0.1 mrad |
| Rotation around radial axis | 0.1 mrad |
| Rotation around normal axis | 16.2 mrad |

**Table 1**: Independent transformations of the primary mirror needed to increase the PSF size to 1 arcmin on axis. For the panel-to-panel alignment, each one follows a Gaussian distribution centered on the ideal position.

| Secondary mirror Global alignment | Value |
|---|---|
| Translation ⊥ to optical axis | 10 mm |
| Translation ∥ to optical axis | 5 mm |
| Tilt | 0.15 mrad |
| Panel alignment | Standard deviation |
| Translation ⊥ to optical axis | 1.1 mm |
| Translation ∥ to optical axis | 4 mm |
| Rotation around tangent axis | 0.2 mrad |
| Rotation around radial axis | 0.3 mrad |
| Rotation around normal axis | 118 mrad |

**Table 2**: Independent transformations of the secondary mirror needed to increase the PSF size to 1 arcmin on axis. For the panel-to-panel alignment, each one follows a Gaussian distribution centered on the ideal position.

## 5 Development of the alignment system

Because of the aspheric optical system and the alignment tolerances, the alignment of the mirror panels requires automated edge sensors and actuators. These tolerances are roughly equivalent to a sub-mm radio telescope operating in the range $100 - 20 \mu m$. Although these requirements are three to four orders of magnitude above the usual diffraction limit of optical telescopes, they are far more demanding than those of current IACT optical systems, such as HESS, MAGIS and VERITAS.

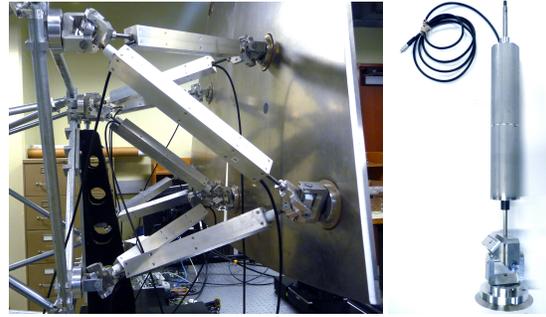

**Fig. 3**: Left: previous prototype of Stewart platform holding and aligning the mirrors on the optical support structure. Right: final prototype of actuator housing and joint.

### 5.1 Alignment system architecture

The mechanical structure of the SCT is developed to be stiff enough to avoid a fully active optic when changing the telescope elevation. It is planned to realign the mirrors only one time per season in order to compensate seasonal thermal deformations.

The alignment system must provide a precise relative alignment of the mirror panels with respect to each other as it is the most constrained specification, as well as a global alignment, capable of aligning the mirrors and the focal plane on the optical axis of the telescope.

The large size ($\sim$1 m$^2$) and small number of mirror panels (72), compared to the traditional IACTs, allow us to mount each of them on a dedicated Stewart platform made of 6 linear actuators, as shown Figure 3 (left). The relative positions of the adjacent panels are monitored by a series of edge sensors located on their edges. Both actuators and edge sensors are connected to a controller board allocated to each panel and mounted behind each of them. During the alignment process, a micro computer on the controller board analyzes and records the relative position of the panels through the edge sensors, then communicates this position to a master computer to calculate the corrected location of this panel, and sends the corresponding commands to the actuators.

### 5.2 Stewart platform

The position of a mirror panel is controlled by six linear actuators arranged in Stewart platform, as presented in Figure 3 (left). These actuators are capable of stepping in increment of 1.5 $\mu$m in a range of 63 mm, while their position is measured by a magnetic encoder. The housing of the actuator is made of an aluminum cylinder with both ends attached to an universal joint, allowing five degrees of freedom without hysteresis in the panel position, larger than several microns.

The actuator and joint shown in Figure 3 (right) are the final versions of a series of prototypes developed at UCLA to optimize their cost, resistance to environment and simplification of the assembly process.

### 5.3 Edge sensors

The mirror panel edge sensor (MPES) units consist of a photosensor and a corresponding light source, each one attached to the edges of adjacent mirror panels and thus presenting independent housings. The purpose of this interface is to measure relative displacements between mirror panels as an input for the panel-to-panel alignment system. The MPESs are required to provide a positional resolution of few $\mu$m over an operational area of approximately 10mm by 10mm.

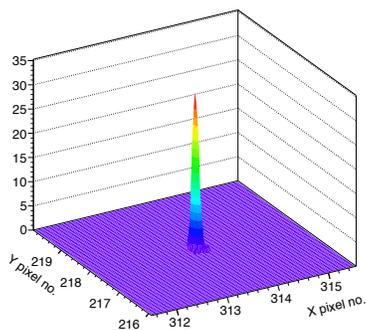

**Fig. 4**: Histogram showing the distribution of image centroids, in pixel coordinates, from 300 different frames. The projected histograms in the X and Y axis are well fitted by a Gaussian function of $\sigma < 0.1$ pixel in both cases. With a screen scale of $48\mu$m/pixel, the individual MPES positional resolution translates to $< 2\mu$m in each axis.

The chosen solution for the photosensor and light source are respectively an economic USB webcam and a laser diode module. Each MPES presents a single optical axis defined by the laser beam, which is in turn orthogonal to the webcam sensor plane. Each of these two main components will present independent housings with additional optical elements: i) the light source housing will also include a collimator to reduce the laser beam width and a filter to reduce its brightness, ii) the sensor housing will include an opal glass screen in front of the sensor that will diffuse the laser beam and create a source at a constant distance from the detector. Such an implementation has been proven to provide relative displacement measurements good to a few microns.

Several USB webcams have been tested in the laboratory, all providing similar positional resolution performances. Figure 4 illustrates the positional resolution obtained in the laboratory for a bare-bone MPES prototype and the present default USB webcam model. The stability of the positional resolution with respect to the position of the laser spot across the MPES operational area has also been proven. Extreme temperatures and long-term usage stress-tests on all MPES electronic components are currently ongoing.

Adjacent mirror panels within the same ring segment will be interfaced with a triad of MPESs, with orthogonal optical axes among themselves, consequently optimizing the measurement of the displacements in the 3 dimensional space. Adjacent mirror panels belonging to different rings will share a single MPES, thus allowing for inter-ring displacement measurements. Additionally, a number of individual MPESs will interface the inner (outer) edge of the inner (outer) ring with the optical support structure (see Fig. 5).

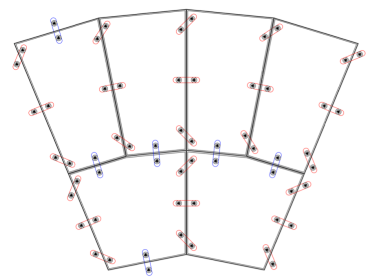

**Fig. 5**: Schematic view of an eighth sector of the SCT primary mirror, showing six mirror panels, as well as the tentative positions of the mounting pads for the MPESs. A number of MPESs will also interface both the outer and inner ring with the optical support structure. Areas defined by a red line indicate the bases of the volumes the MPESs configured in triads will reside in. Areas defined by a blue line indicate the bases of the volumes the single MPESs will reside in.

### 5.4 Controller board

Each panel of the primary and secondary mirrors will hold a controller board behind it, responsible for the collection of the signal coming form the edge sensors and magnetic encoders, the signal processing using a micro computer, and the transmission of the commands to the actuators.

A working prototype of such board is already available, with the capability to precisely control six actuators and analyze the images coming from 7 edge sensors. This prototype is currently being revised to integrate USB and ethernet interface protocols.

## Summary

During the first year of the development of the SCT prototype, the conceptual designs of the main elements of the optical system have been defined, and the prototypes of some of these elements have been fabricated. The construction of the SCT prototype will start in 2015 at the Fred Lawrence Whipple Observatory, in southern Arizona.

## Acknowledgements
This research is supported by the US National Science Foundation under the Major Research Instrumentation grant no MRI-Phy-1229792.

We also gratefully acknowledge support from the agencies and organizations listed in this page: http://www.cta-observatory.org/?q=node/22.